\documentclass{aastex631}
\usepackage[T1]{fontenc}
\shorttitle{A Failed Eruption Involving External Reconnection}
\shortauthors{Chen et al.}

\usepackage{CJK}
\usepackage{booktabs}
\usepackage{threeparttable}
\graphicspath{{./}{figure/}}

\begin{document}
	\begin{CJK*}{UTF8}{gbsn}
		\title{Observations of a Failed Solar Filament Eruption Involving External Reconnection}

		\author[0000-0003-1220-1582]{Yuehong~Chen({\CJKfamily{gbsn}陈悦虹})}
		\affiliation{School of Astronomy and Space Science, Nanjing University, Nanjing 210023, People's Republic of China; xincheng@nju.edu.cn and el2718chenjun@nju.edu.cn}
		\affiliation{Key Laboratory of Modern Astronomy and Astrophysics (Nanjing University), Ministry of Education, Nanjing 210023, People's Republic of China}
		
		\author[0000-0003-2837-7136]{Xin~Cheng({\CJKfamily{gbsn}程鑫})}
		\affiliation{School of Astronomy and Space Science, Nanjing University, Nanjing 210023, People's Republic of China; xincheng@nju.edu.cn and el2718chenjun@nju.edu.cn}
		\affiliation{Key Laboratory of Modern Astronomy and Astrophysics (Nanjing University), Ministry of Education, Nanjing 210023, People's Republic of China}

		\author[0000-0003-3060-0480]{Jun~Chen({\CJKfamily{gbsn}陈俊})}
		\affiliation{School of Astronomy and Space Science, Nanjing University, Nanjing 210023, People's Republic of China; xincheng@nju.edu.cn and el2718chenjun@nju.edu.cn}
		\affiliation{Key Laboratory of Modern Astronomy and Astrophysics (Nanjing University), Ministry of Education, Nanjing 210023, People's Republic of China}
            \affiliation{Purple Mountain Observatory, Chinese Academy of Sciences, Nanjing 210023, China}
		
		\author[0000-0001-9856-2770]{Yu~Dai({\CJKfamily{gbsn}戴煜})}
		\affiliation{School of Astronomy and Space Science, Nanjing University, Nanjing 210023, People's Republic of China; xincheng@nju.edu.cn and el2718chenjun@nju.edu.cn}
		\affiliation{Key Laboratory of Modern Astronomy and Astrophysics (Nanjing University), Ministry of Education, Nanjing 210023, People's Republic of China}
		
		\author[0000-0002-4978-4972]{Mingde Ding({\CJKfamily{gbsn}丁明德})}
		\affiliation{School of Astronomy and Space Science, Nanjing University, Nanjing 210023, People's Republic of China; xincheng@nju.edu.cn and el2718chenjun@nju.edu.cn}
		\affiliation{Key Laboratory of Modern Astronomy and Astrophysics (Nanjing University), Ministry of Education, Nanjing 210023, People's Republic of China}
		
		\begin{abstract}
			We report a failed solar filament eruption that \edit2{involves} external magnetic reconnection in a quadrupolar magnetic configuration. The evolution exhibits three kinematic evolution phases: a slow-rise phase, an acceleration phase, and a deceleration phase. In the early slow rise, extreme-ultraviolet (EUV) brightenings appear at the expected null point above the filament and are connected to the outer polarities by the hot loops, indicating the occurrence of a breakout reconnection. Subsequently, the filament is accelerated outward, accompanied by the formation of low-lying high-temperature post-flare loops ($>$ 15 MK), complying with the standard flare model. However, after 2--3 minutes, the erupting filament starts to decelerate and is finally confined in the corona. The important finding is that the confinement is closely related to an external reconnection as evidenced by the formation of high-lying large-scale hot loops ($>$ 10 MK) with their brightened footpoints at the outer polarities, the filament fragmentation and subsequent falling along the newly formed large-scale loops, as well as a hard X-ray source close to one of the outer footpoint brightenings. We propose that, even though the initial breakout reconnection and subsequent flare reconnection commence and accelerate the filament eruption, the following external reconnection between the erupting flux rope and overlying field, as driven by the upward filament eruption, makes the eruption finally failed, as validated by the numerical simulation of a failed flux rope eruption.

		\end{abstract}
		
		
		\section{introduction} \label{sec:intro}
		The solar atmosphere is highly dynamic and frequently produces explosive events \citep{low1996}, such as flares, filament/prominence eruptions, and coronal mass ejections (CMEs). These phenomena can expel significant amounts of high-speed plasma into the interplanetary space, accompanied by high-energy particles and high-intensity electromagnetic radiation, which may pose severe impacts on the safety of the terrestrial environment and human high-tech activities \citep{schwenn2006}.
		
		Among all solar explosive events, the eruption of filaments \edit2{is perhaps the most intriguing one}. Based on the amount of ejected mass and escaping magnetic structures, \citet{gilbert2007} have classified filament eruptions into three types: full, partial, and failed eruptions. Failed eruptions generally exhibit filament mass and magnetic structures that abruptly decelerate and stop at a certain height after an initial eruptive-like acceleration \citep[e.g.,][]{Ji2003}. The mass eventually drains back into the solar surface, and thus no CME is observed in white-light coronagraph images. 
        
		With multiple wavelength observations, the main characteristics of failed filament eruptions have been disclosed such as the rotation of the entire magnetic configuration \citep{song2018}, falling back of associated materials along original loops \citep{kuridze2013,zhang2021}, as well as co-spatial hard X-ray (HXR) and/or microwave (MW) sources with confined plasmoids \citep{alexander2006,netzel2012,song2014,kushwaha2015,cheng2018}. Interestingly, \citet{chen2013} found that the erupting materials seem to be confined by the overlying arcades that initially expand for a while but are quickly stopped. In addition, the oscillation of the overlying loops could be triggered by the failed eruption \citep[e.g.,][]{mrozek2011}. 
  
        The failed eruptions sometimes can produce features alike to a successful eruption including the heating of the erupting flux rope \citep[e.g.,][]{li2022} and the separation of two flare ribbons \citep{cheng2015}. In a confined filament eruption, through nonlinear force-free modeling, \citet{guo2010} even found that the induced two ribbons have a good correspondence to the photospheric imprints of large-scale quasi-separatrix layers (QSLs), similar to eruptive eruptions.

		The interpretation of filament eruptions often resorts to ideal magnetohydrodynamic (MHD) instabilities of magnetic flux rope (MFR), which is defined as a set of magnetic field lines winding around a common axis \citep[e.g.,][]{Cheng2017}. The most popular one is torus instability \citep[TI,][]{torok2005,kliem2006}, the occurrence of which requires a sufficiently rapid decrease of the background field with height such that the confining force due to the external field decreases faster than the upward hoop force. The critical value for TI, known as the decay index ($-\partial~{\rm ln} \left | \textbf{B}_{{\rm p}} \right |/\partial~{\rm ln}~h$) is suggested to be between 1.1 and 1.5 \citep{kliem2006,demoulin2010,olmedo2010,zuccarello2015}. A lower decay index than the critical value may explain the failed eruptions \citep{cheng2011a, cheng2020}. Nonetheless, laboratory experiments \edit2{\citep[e.g.,][]{myers2015}} demonstrate that the final fate of an MFR eruption should be determined by the mutual cooperation of kink instability \citep[KI,][]{hood1979, Torok2004} and TI.
		
		Some other causes of failed eruptions have also been proposed in numerical simulations with a focus on the force analysis of an erupting MFR,  such as the tension of the strong guide field component of the overlying field \citep{torok2005,myers2015}, the tension of the strong overlying arcade field \citep{DeVore2008}, and the Lorentz force due to the non-axisymmetry of the MFR \citep{zhong2021}. Some other explanations also highlight the reconnection between the writhed MFR and the overlying arcade \citep{amari1999, hassanin2016}. All these suggest that the external magnetic field has an important influence on the final fate of the erupting MFR.

		Recently, we proposed a new model for failed MFR eruptions that originate from the quadrupolar configuration, where the inner bipole and the outer bipole are antiparallel so that hyperbolic flux tubes (HFTs) exist both below and above the MFR \citep{chen2023}. \edit2{We defined two types of magnetic reconnection occurring at the higher HFT above the MFR. The first one is the breakout reconnection between the flux from the outer polarities and the flux above the flux rope from the inner polarities. The second one is the external reconnection between the flux of the outer polarities and the flux of the MFR itself. Additionally, the reconnection at the lower HFT below the MFR is referred to as to the ``lower reconnection'', which is in principal the same as the current sheet reconnection in the standard flare model.} It was found that the initial breakout reconnection promotes the MFR eruption with a combination of the lower reconnection, which agrees with the breakout model, as found in both observations \citep{aulanier2000,mandrini2006,shen2012,chen2016,kumar2018} and simulations \citep{antiochos1999,macneice2004,lynch2008,karpen2012,wyper2018}. Nonetheless, the external reconnection at the later stage significantly erodes the poloidal flux of the MFR, decreases the upward Lorentz force, and finally results in a failed eruption. 

        The external reconnection between the erupting MFR and ambient fields has been noticed previously in both simulations \citep{cohen2010,aulanier2019} and observations \citep{chintzoglou2017,dudik2019,ding2022}. It was only found to play an important role in the drifting of the MFR footpoints \citep{xing2020,zhong2021}.
  
	   In this manuscript, we investigated a failed filament eruption in detail and find that its failure is most likely caused by the external reconnection after carefully comparing the observations and the numerical simulation of the failed MFR eruption by \citet{chen2023}. For the rest of the paper, we present observations and methods in Section \ref{sec2}. The results are displayed in Section \ref{sec3}, which is followed by a summary and discussion in Section \ref{sec4}.

		\section{Observations}\label{sec2}
		The X1.1 flare under study occurred in the AR NOAA 13006 on 2022 May 3 and \edit2{was related to} a failed filament eruption. This event was well observed by the Atmospheric Imaging Assembly \citep[AIA,][]{lemen2012} on board \emph{Solar Dynamics Observatory} \citep[\emph{SDO},][]{pesnell2012}. Based on the observed intensities at the six coronal passbands of AIA, we adapted the sparse algorithm \citep{cheung2015,su2018} to perform differential emission measure (DEM) inversion over a temperature range of $\rm{log}~\emph{T/K}\in [5.5,7.6]$ and a grid spacing of $\Delta~\rm{log}~\emph{T} = 0.05$. The robustness of the DEM inversion was validated through 200 Monte Carlo simulations. With $n$ = 43 temperature bins, the DEM-weighted average temperature $\left \langle {T} \right \rangle$ was defined as follows:
		\begin{equation}
			\left \langle {T} \right \rangle = \frac{\sum\limits ^{{n}}_{{i}}\mathrm{DEM}_{{i}}~{T}_{{i}}\bigtriangleup{T_{i}}}{\sum\limits ^{{n}}_{{i}}\mathrm{DEM}_{{i}}\bigtriangleup{T_{i}}}.
		\end{equation}

		We also examined the observations of the Extreme Ultraviolet Imager \citep[EUVI,][]{wuelser2004} on board \emph{Solar Terrestrial Relations Observatory-Ahead} \citep[\emph{STEREO-A},][]{kaiser2008}, which was separated from the Earth by -31$^{\circ}$ in heliographic longitude during the event, and of the Extreme Ultraviolet Imager \citep[EUI,][]{rochus2020} on board \emph{Solar Orbiter} \citep{muller2020}, which was about 163$^{\circ}$ ahead of the Earth in heliographic longitude and was located at 0.7 AU away from the Sun on 2022 May 3. \emph{Solar Orbiter}/EUI consists of three telescopes, including two High Resolution Imagers (HRIs) and one Full Sun Imager (FSI). For this event of interest, it was only observed by the FSI, which provides images from \edit2{another perspective} at the 174 {\AA} and 304 {\AA} passbands with a pixel scale of around 2.3 Mm, and time cadences of 10 minutes and 30 minutes, respectively. 
  
        We then investigated the HXR data provided by the Spectrometer Telescope for Imaging X-rays \citep[STIX,][]{krucker2020} on board \emph{Solar Orbiter}. The HXR sources were reconstructed using the Expectation Maximization \citep[EM,][]{massa2019} algorithm. \edit2{The HXR data from Gamma-ray Burst Monitor \citep[GBM,][]{meegan2009} on board \emph{Fermi Gamma-ray Space Telescope} (\emph{Fermi}) were also used}, whereas the 1--8 {\AA} soft X-ray (SXR) flux data were obtained from \emph{Geostationary Operational Environmental Satellite} (\emph{GOES}). Note that the times of all detectors have been converted to the times at the Earth.

		We extrapolated the three-dimensional (3D) coronal potential magnetic field using a Green's function method \citep{chiu1977}, based on the radial magnetogram, which is directly taken from Space-weather HMI Active Region Patches \citep[SHARPs,][]{bobra2014} from the Helioseismic and Magnetic Imager \citep[HMI,][]{schou2012} on board \emph{SDO}. Because the event under study took place at the solar limb, we used a SHARP magnetogram taken three days later (13:00 UT on 2022 May 6) when the hosting active region (AR) had rotated onto the disk. Doing so did not influence our results qualitatively because the distribution of the dominant magnetic flux for the source region generally evolved little during the three days and thus the corresponding potential field structure was thought not to change significantly \citep[also see][]{cheng2020}.
		
		\section{Results}\label{sec3}
		
		\subsection{Overview of Observations}\label{sec31}
\begin{figure}
			\epsscale{1.0}
			\plotone{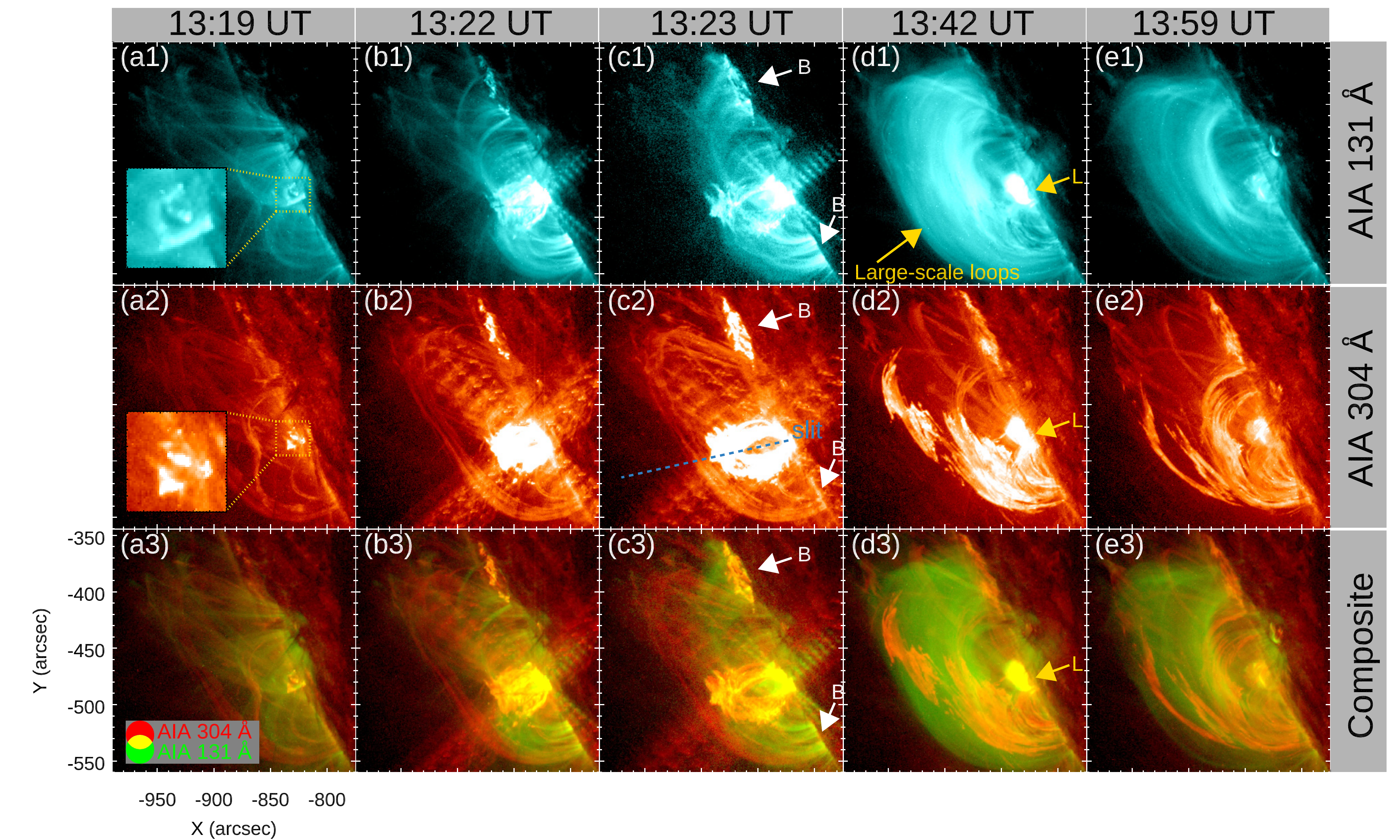}
			\caption{AIA images showing the failed filament eruption. The top (middle) panels are the AIA 131 (304) {\AA} images. The bottom panels are composite of the AIA 131 {\AA} (green) and 304 {\AA} (red) images. The sub-panels indicated by the yellow boxes in panels (a1) and (a2) are the zoom-in of the filament. The white arrows denote the remote brightnesses (B). The yellow arrows denote the low-lying post-flare loops (L) and overlying large-scale loops. The blue dashed line in panel (c2) denotes the slit along the eruption direction. An animation is available online to show the eruption during 13:00--14:47 UT with AIA 131 {\AA} images, AIA 304 {\AA} images, the composite of AIA 131 {\AA} and 304  {\AA} images, and the DEM-weighted average temperature images. The duration of the animation is 15 s.}
			\label{fig1}
		\end{figure}

		Prior to the eruption, a filament was observed to be located in the lower corona as shown in the inset of Figure \ref{fig1}(a1)--(a2). At approximately 13:16 UT, some weak extreme-ultraviolet (EUV) brightenings appeared around the filament. At the same time, the filament experienced a slow rise. This slow rise persisted for about 4 minutes until the flare started. Afterward, the filament was accelerated outward as shown in Figure \ref{fig1}(b), producing remote brightenings (``B'' in Figure \ref{fig1}(c)). Subsequently, some small-scale low-lying loops (``L'' in Figure \ref{fig1}(d)) appeared underneath the erupting filament. After a short period of acceleration, the rising filament started to decelerate, became fragmented, and then fell down to the positions away from the flare region. The falling of filament materials seems to be along the newly formed large-scale hot loops appearing at the AIA 131 {\AA} passband~as displayed in Figures \ref{fig1}(d)--(e). More details on the process of the failed filament eruption can be seen in the accompanying animation of Figure \ref{fig1}.
		
\begin{figure}
			\epsscale{0.8}
			\plotone{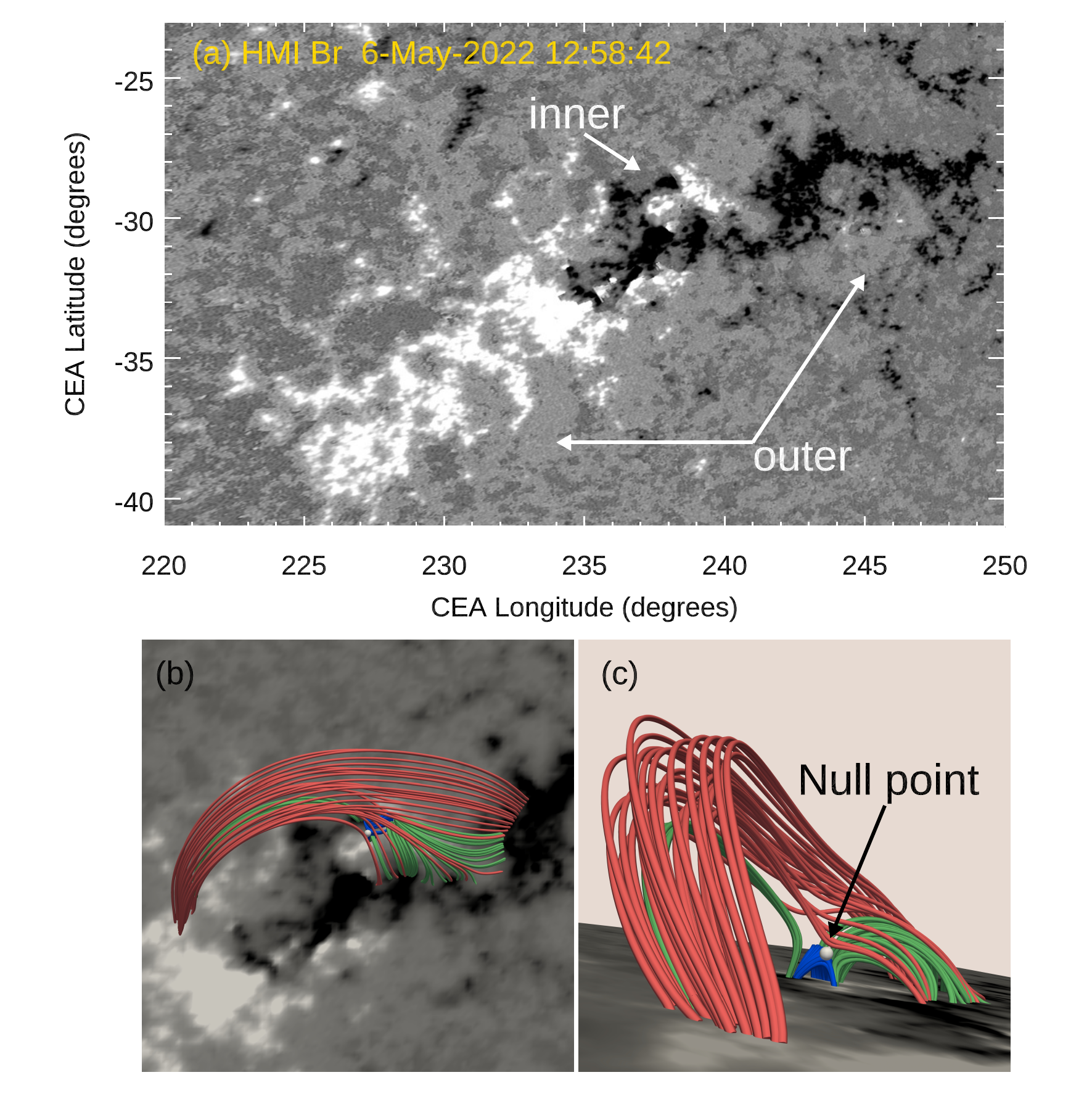}
			\caption{(a) The radial component of the Lambert Cylindrical Equal-Area (CEA) vector magnetic field from SHARP with a linear scale saturating at ±500 G and the white (black) region is positive (negative). The white arrows denote the inner and the outer polarities. (b)--(c) The extrapolated 3D potential field structure from the side and top views, the center-lobe flux, side-lobe flux~and high-lying flux~are in blue, green, and red color, respectively. The gray ball indicates the position of the null point.}
			\label{fig2}
		\end{figure}

		The distribution of the vertical magnetic field $B_r$ of the associated active region at the photosphere (Figure \ref{fig2}(a)) indicates quadrupolar polarities including two smaller inner polarities and two larger outer polarities. The extrapolated 3D potential field (Figure \ref{fig2}(b)--(c)) confirms a magnetic field configuration that consists of four flux systems: a center-lobe flux (colored in blue), two side-lobe fluxes (in green), and a large-scale high-lying flux (in red), consistent with the quadrupolar configuration presented in \citet{antiochos1999}. \edit2{In the extrapolation, a null point exists between the center-lobe flux and large-scale high-lying flux. During the eruption, the expected null point} may evolve toward the high-lying HFT where the breakout reconnection occurs. Note that, due to the limitation of the potential field model, we did not get the twisted flux rope and the HFT below it in the extrapolated field as in \citet{chen2023}. \edit2{The existence of the flux rope is however indicated by the observed filament.}

		To investigate the kinematic evolution of the failed filament eruption, we inserted an artificial slit along the eruption direction as shown in Figure \ref{fig1}(c2) and created a distance-time (DT) plot as shown in Figure \ref{fig3}(a). We estimated the velocities of the leading front of the erupting filament from the height-time measurements using the IDL routine \texttt{deriv.pro}. Figure \ref{fig3}(b) shows that the filament underwent a three-phase kinematic evolution: a slow-rise phase, an acceleration phase, and a deceleration phase. They are analyzed in detail in the following subsections.
		
		\subsection{The Slow-rise Phase}\label{sec32}
		The filament rose slowly before 13:19:53 UT with a velocity of around 10 km~s$^{-1}$, which was accompanied by \edit2{weak} and continuous EUV brightenings around the filament (the red arrows in Figures \ref{fig4}(a)--(b)), indicating the activation of the eruption.
		
		Besides, as shown in Figures \ref{fig4}(a)--(b), EUV brightenings also appeared in the corona with a height of about 30 Mm. They were connected to the outer polarities by the slightly bright side-lobe loops (``SLs'' in Figure \ref{fig4}(a)). Because these EUV brightenings occurred above the filament, we speculate that a magnetic reconnection pertinent to the high-lying null point as indicated by the extrapolated 3D structure (Figure \ref{fig2}(b)--(c)) may take place. It could be similar to the breakout reconnection between the center-lobe flux above the filament and the large-scale high-lying flux, generating new side-lobe loops corresponding to those we observed. In this event, we only observed the southern side-lobe loops whereas the northern loops were ambiguous, which might be due to the asymmetry of the magnetic topology. It has been shown that the breakout reconnection can help remove the constraint of the overlying field and thus facilitate the filament eruption \citep{macneice2004,lynch2008,karpen2012,chen2016}.

		\subsection{The Acceleration Phase}\label{sec33}
		From 13:19:53 UT to 13:22:17 UT, the filament exhibited a rapid outward acceleration from around 10 km~s$^{-1}$ to 375 km~s$^{-1}$, which \edit2{was} accompanied by a great increase in EUV emissions beneath the filament, as displayed in Figures \ref{fig4}(c)--(d). Afterward, some short bright post-flare loops appeared (Figure \ref{fig4}(e)). The DEM analysis (Figure \ref{fig4}(f)) shows that the post-flare loops are heated to a DEM-weighted average temperature of above 15 MK. \edit2{According to the standard flare model}, the largely brightened post-flare loops are a consequence of the lower flare reconnection, which adds poloidal flux to the filament-hosting MFR at the same time \citep{cheng2011b,xing2020}. As a result, the whole MFR system impulsively grows and erupts.

		Interestingly, it was found that some bright blobs appeared repeatedly and moved along the bright large-scale loops. We speculate that they may be related to the plasmoids generated in the high-lying current sheet, which can be formed through the collapse of the breakout null point during the main acceleration phase, as simulated by \citet{karpen2012} and \citet{wyper2018}. In 3D, the plasmoids are tiny flux ropes formed by tearing mode instability \citep{wang2023}. However, the confirmation of the blobs being plasmoids is restricted by the resolution of the current observations. The accompanying animation of Figure \ref{fig4} provides more details.
		
		\subsection{The Deceleration Phase}\label{sec34}
\begin{figure}
			\epsscale{0.8}
			\plotone{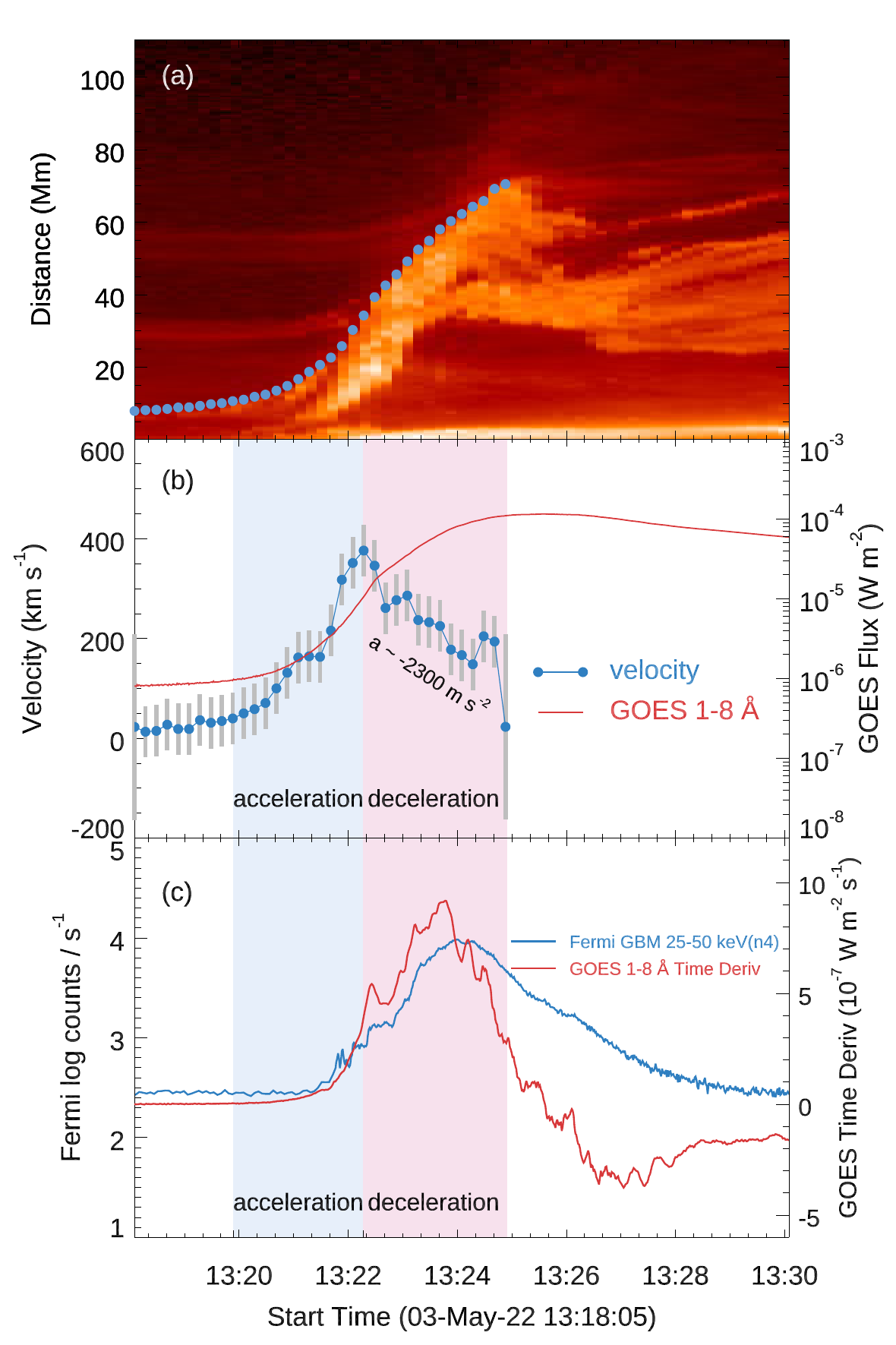}
			\caption{(a) DT plot of the AIA 304 {\AA} images along the slit in Figure \ref{fig1}(c2), the blue circles denote the height-time measurements of the filament leading front. (b) The \emph{GOES} 1--8 {\AA} SXR flux and the temporal evolution of the filament velocity, with the errors from the uncertainty in height measurements ($\sim$2 pixels, 877 km). (c) The HXR flux of the \emph{Fermi} 25--50 keV channel and the time derivative of the GOES SXR flux. A 10-second smoothing is applied to the latter. The blue and red background in panels (b)--(c) indicate the acceleration phase and deceleration phase, respectively.}
			\label{fig3}
		\end{figure}

\begin{figure}
			\epsscale{0.7}
			\plotone{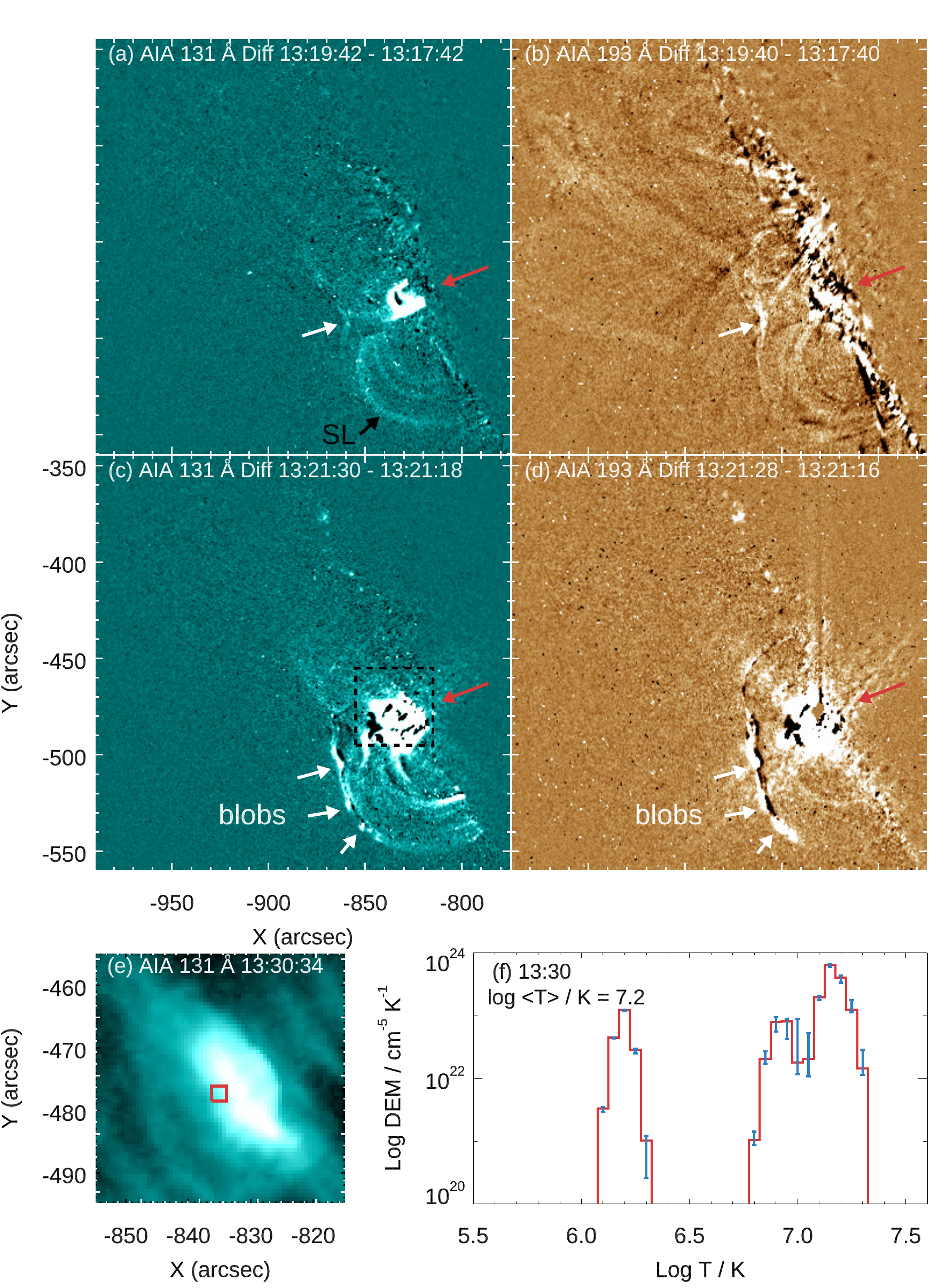}
			\caption{The AIA 131 {\AA} and 193 {\AA} running-difference images showing the slow-rise phase (a)--(b) and the acceleration phase (c)--(d) of the eruption. The red, black, and white arrows denote the EUV brightenings beneath the filament, the side-lobe loops (SL), and EUV brightenings or blobs, respectively. (e) The AIA 131 {\AA} image displaying the low-lying post-flare loops as outlined by the box in panel (c). (f) DEM distribution of a small region at the post-flare loops as shown by the red box in panel (e). The red histogram is the result directly from observational data. The blue error bars are uncertainties, demonstrating the upper and lower limits, in which 90$\%$ of Monte Carlo solutions larger and smaller than the best fitting values based on observational data are contained, respectively. An animation is available online to show the running difference images at AIA 131 {\AA} and 193 {\AA} passbands during 13:15--13:29 UT. The duration of the animation is 5 seconds.}
			\label{fig4}
		\end{figure}

		After 13:22:17 UT, the filament entered the deceleration phase. As shown by Figure \ref{fig3}(b), the maximum deceleration is about 8 times greater than the solar gravitational constant ($g$ = 274~m s$^{-2}$), suggesting a stronger downward force exerting on the filament besides the gravity. 
		
		An interesting finding is that the \emph{GOES} 1--8 {\AA} SXR emission continuously increases during the deceleration phase. The peak of SXR emission is delayed by about 3 minutes to that of the filament velocity, which is a common phenomenon for failed eruptions \citep{huang2020} but different from successful eruptions, in which the SXR emission and CME velocity almost reach their maxima simultaneously \citep[e.g.,][]{zhang2001}. A possible reason is that the reconnection proceeds even though the filament has decelerated so that the SXR emission continuously increases during the deceleration phase. This is well confirmed by the fact that the \emph{Fermi} 25--50 keV HXR flux and the time derivative of the \emph{GOES} SXR 1--8 {\AA} flux both peak during the deceleration phase (Figure \ref{fig3}(c)). One may speculate that the abnormally strong HXR emission in the deceleration phase is due to that the majority of HXR emissions in the acceleration phase, likely located at the footpoints of the low-lying post-flare loops, are blocked by the solar limb. If the postulated scenario holds true and the Neupert effect \citep{Neupert1968} is valid, we should observe a stronger component in the time derivative of the SXR emission in the acceleration phase because, obviously, their sources, i.e., post-flare loops, are not obstructed. However, the maximum does appear in the deceleration phase (Figure \ref{fig3}(c)), which tends to support a stronger reconnection than that in the acceleration phase.
		
\begin{figure}
			\epsscale{0.7}
			\plotone{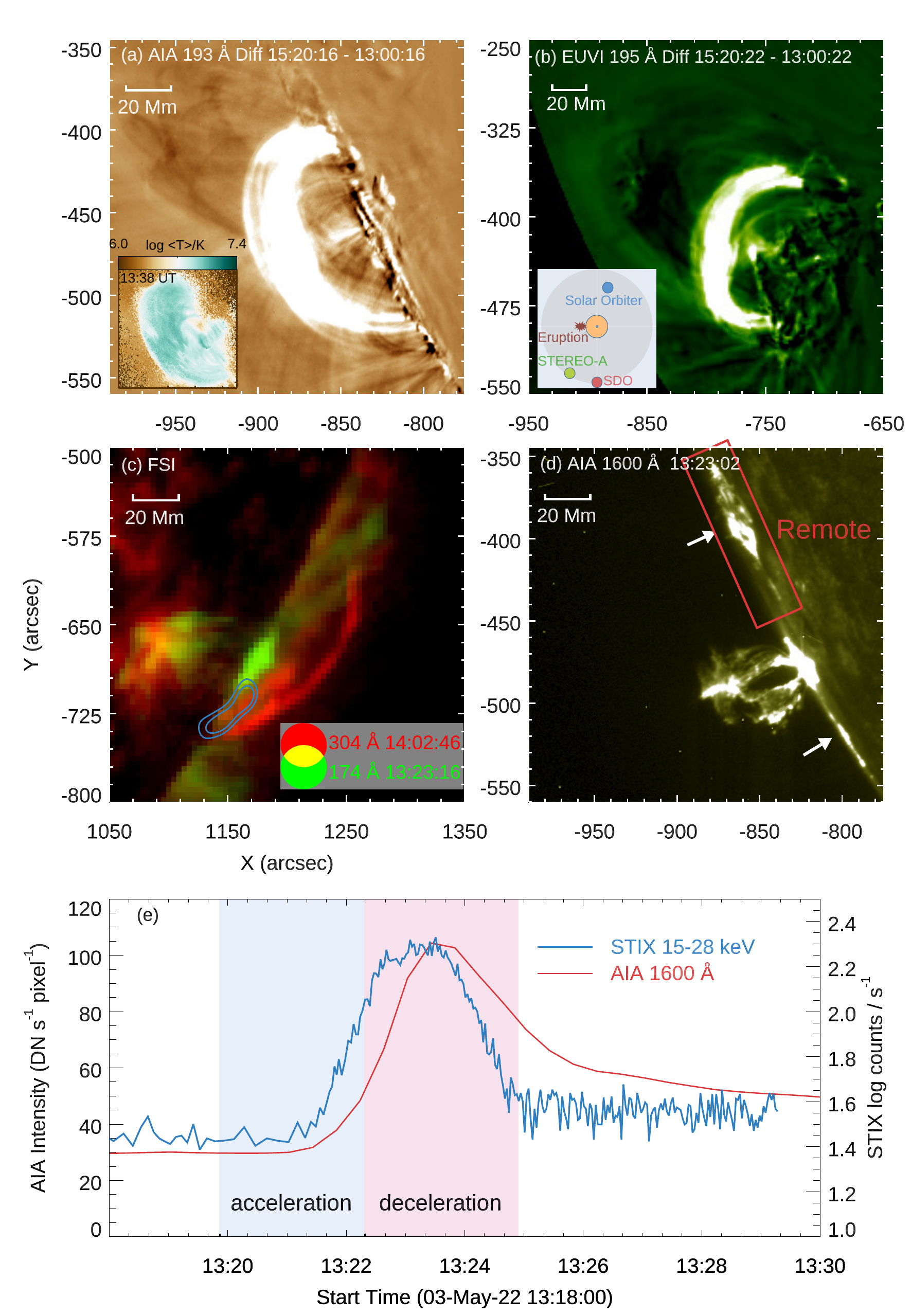}
			\caption{(a)--(b) The AIA 193 {\AA} and EUVI 195 {\AA} base-difference images showing the newly formed large-scale loops. The DEM-weighted average temperature map is inserted in panel (a). The relative locations of the Sun, \emph{SDO}, \emph{STEREO-A}, and \emph{Solar Orbiter} are shown in panel (b). (c) Composite of the FSI 304 {\AA} (red) and 174 {\AA} (green) images and the 50$\%$ and 70$\%$ contours of the STIX HXR emission maximum are overplotted in blue. (d) The AIA 1600 {\AA} image displaying two remote brightenings pointed out by two white arrows. The red box outlines the region of northern remote brightening. (e) The STIX 15--28 keV lightcurve and the AIA 1600 {\AA} pixel-average intensity over the box in panel (d). The backgrounds in different colors are the same as Figure \ref{fig3}. An animation is available online to show the eruption during 13:00--16:00 UT with \emph{SDO}/AIA 193 {\AA} and \emph{STEREO}/EUVI 195 {\AA} images. The duration of the animation is 9 s.}
			\label{fig5}
		        \end{figure}

		The map of DEM-weighted average temperature (the inset of Figure \ref{fig5}(a)) demonstrates that the large-scale loops formed during the deceleration phase were heated up to $\sim$10 MK. These loops cooled down to around 1.2 MK after 2 hours of cooling, and were thus visible in AIA 193 {\AA} and EUVI 195 {\AA} bands as shown in Figures \ref{fig5}(a)--(b). The footpoints of the large-scale loops corresponded to the outer brightenings. Besides, the filament became fragmented and the filament material was transferred to the positions where the large-scale loops were newly formed, as shown by the spatial stereoscopic perspective via both \emph{SDO}/AIA and \emph{STEREO}/EUVI (see the animation of Figure \ref{fig5}). Generally, the filament materials are believed to be suspended at magnetic dips of a coherent MFR in the static phase, and they are expected to return to the footpoints of the MFR in a failed eruption if the flux of the MFR does not reconnect with the ambient field. This obviously differs from what was observed for this event. We thus argue that the external reconnection between the filament-associated MFR and the large-scale overlying field most likely takes place so that the fragmented materials can be transferred to new positions.
   
		In order to determine the location of the HXR emission during the deceleration phase, we reconstructed the HXR 15--28 keV sources with \emph{Solar Orbiter}/STIX data in the time interval from 13:22:19 UT to 13:24:53 UT. They were plotted over the \emph{Solar Orbiter}/FSI 174 {\AA} and 304 {\AA} images in Figure \ref{fig5}(c). The result indicates that the HXR 15--28 keV emission mainly came from the region of the southern EUV/UV brightening that is closely related to the external reconnection. The magnetic field has a northwest-southeast direction, explaining the reason why we only detected the southern HXR source from the the point of view of \emph{Solar Orbiter}. Furthermore, the STIX 15--28 keV emission sustains a high level for about 1 minute after its peak during the deceleration phase (Figure \ref{fig5}(e)), and the temporal evolution of the AIA 1600 {\AA} intensity over the northern remote brightening (enclosed by the red box in Figure \ref{fig5}(d)) is similar to that of the HXR 15--28 keV emission. These results indicate that the external reconnection is also fairly strong during the deceleration phase.

            \subsection{Magnetohydrodynamic Simulation of External Reconnection}\label{sec35}
\begin{figure}
			\epsscale{1.0}
			\plotone{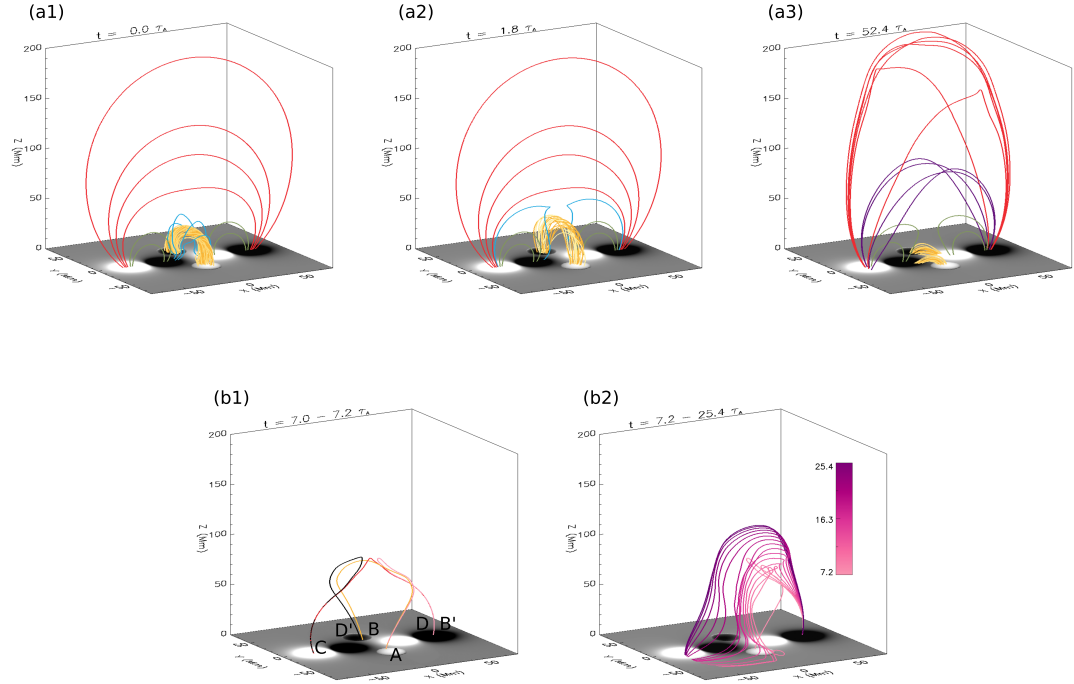}
			\caption{Snapshots of 3D magnetic field lines from the simulation in \citet{chen2023} of a failed MFR eruption in quadrupolar configuration. The bottom boundaries demonstrate the distribution of the $B_z$ component of the magnetic field. (a1) The initial magnetic field configuration with different colors denoting different magnetic fields, the red for the high-lying field, the blue (green) for the center (side) lobe, and the yellow for the MFR. (a2) The rending of the breakout reconnection. The blue lines represent the newly formed side lobe. (a3) The rending of the external reconnection, by which the MFR is completely eroded. The formation of the new large-scale overlying loops (the purple lines) is explained in panels (b1)--(b2). In panel (b1), the yellow (pink) field line is traced from footpoint A at t = 7.0 $\tau_{\rm A}$ (7.2 $\tau_{\rm A}$), suggesting that the reconnection changes AB to AB'. The red (black) field line is traced from footpoint C at t = 7.0 $\tau_{\rm A}$ (7.2 $\tau_{\rm A}$), suggesting that the reconnection changes CD to CD'. In panel (b2), the lines from pink to purple colors denote a successive reconnection process that forms the large-scale field line as illustrated by its footpoint migration. The accompanying animation provides the evolution of the magnetic field lines during $t \in [0, 55.3]~\tau _{A}$. The duration of the animation is 11 seconds.}
			\label{fig6}
		\end{figure}

		To understand the role of such an external reconnection~between the erupting MFR and the outer-bipole-related high-lying field, we analyzed the MHD numerical simulation data in \citet{chen2023} and demonstrated the transition from the breakout to external reconnection in Figure \ref{fig6} \edit2{and an accompanying animation}.
  
		Before the eruption, the MFR is settled below the inner center-lobe flux~(Figure \ref{fig6}(a1)). The breakout process above the MFR transfers the center-lobe flux~to the side lobes (Figure \ref{fig6}(a2)), which is observed as the side-lobe loops in Figure \ref{fig4}(a). 
  
        When the center-lobe flux~above the MFR is exhausted, the MFR subsequently reconnects with the high-lying arcades. In this way, the MFR is eroded, which can be also imagined as peeling an onion for simplicity. In the end, the MFR is completely eroded, generating a great amount of large-scale flux in the high corona (the purple lines in Figure \ref{fig6}(a3)). We suggest that the newly formed field lines correspond to the heated large-scale loops observed in the current event (Figures \ref{fig5}(a)--(b)). 
  
        Figures \ref{fig6}(b1)--(b2) display more details of the formation process of these large-scale loops. Firstly, as shown by Figure \ref{fig6}(b1), the field line from the MFR (AB) reconnects with the pre-existing high-lying field~line (CD) and generates new field lines (AB' and CD'). After one footpoint of the line from the MFR is changed to the outer bipole (i.e., AB is changed to AB'), the other footpoint (A) drifts (Figure \ref{fig6}(b2)), as driven by the continuous external reconnection. The final consequence is that the newly reconnected magnetic field lines gradually migrate from the inner to outer polarities. The two outer footpoints of the newly formed lines in the simulation correspond to the footpoints of the large-scale loops in observations. They are also visible at the AIA 1600 {\AA} passband (Figure \ref{fig5}(d)), indicating that the heating from the external reconnection reaches the lower atmosphere where the 1600 {\AA} emission is formed.  
        
	Different from the aerodynamic drag effect proposed in previous studies \citep[e.g.,][]{vrsnak2004,shi2015}, the external reconnection is believed to be the main reason for the deceleration of the filament eruption. As proved in our simulation, the external reconnection is able to decrease the poloidal flux of the erupting MFR and the upward Lorentz force, \edit2{and might thus be regarded as a mechanism leading to failed eruptions.}

		\section{Summary and Discussions}\label{sec4}
		In this paper, we investigate a failed filament eruption involving external reconnection enabled by its quadrupolar magnetic field configuration. The kinematic evolution of the failed eruption has three phases, including the slow-rise phase, acceleration phase, and deceleration phase.
		
		During the early slow-rise phase, the filament ascended with a speed of around 10~km~s$^{-1}$, similar to that for successful eruptions \citep{cheng2020}. The EUV brightenings appearing at the expected null point and being connected to the outer polarities of the active region by the bright side-lobe loops indicate the occurrence of the breakout reconnection. It may play a role in the removal of the constraint from the overlying field \citep{antiochos1999}, thus facilitating the subsequent filament eruption.
		
		During the acceleration phase, \edit2{the filament was impulsively ejected.} The enhanced emissions from the high-temperature post-flare loops beneath the filament suggest the flare reconnection, which heats the flare plasma and accelerates the MFR eruption simultaneously.

		During the filament deceleration, the large-scale high-temperature overlying loops were formed with enhanced emissions at their footpoints, as a result of external reconnection between the filament-associated MFR and the overlying field. At the same time, the filament was split into fragments that then fell down along the newly formed large-scale overlying loops. Due to the external reconnection, the filament eruption decelerated and was finally confined in the corona. \edit2{It is obviously different from the interpretations of the failed eruptions in bipolar fields \citep[e.g.,][]{guo2010,Zhou2019}}.

        The breakout reconnection at the null point in the quadrupolar configuration has been argued to be able to promote the eruption \citep[e.g.,][]{shen2012,chen2016}, complying with the ``breakout'' model \citep{antiochos1999}. However, in our study, after the initial breakout reconnection, the external reconnection likely starts to replace the breakout one and erodes the erupting structure. In the end, it makes the eruption fail by reducing the poloidal flux and thus the Lorentz force of the filament-hosting MFR. This erosion process may also exist in previous events but its role in constraining the eruption has not been stated explicitly \citep[e.g.,][]{chintzoglou2017}. \edit2{It should be addressed that precisely determining the transition time from the breakout to external reconnection is really difficult in both observations and simulations, even though the time interval when one type of reconnection dominates the other can be estimated roughly.}

		It is worth mentioning that the tension force of the large-scale overlying field may also play a role in constraining the eruption \citep[e.g.,][]{DeVore2008}. However, as suggested by \citet{chen2023},  this kind of tension force only works for a while at the very early stage of the eruption and hence can not be the main reason for the failure of the eruption.

        We suggest the strong HXR emissions appearing at the outer southern polarity during the deceleration phase to be a critical discriminator for the external reconnection. The northern source was not observed because of the obstruction of the solar disk. \edit2{However, this source could have existed, as indicated by the intense AIA 1600 {\AA} brightenings at the north footpoints.} Furthermore, it is noteworthy that, for the event under investigation, the strongest HXR emissions appeared in the deceleration phase, in contrast to earlier studies that show stronger HXR emissions in the acceleration phase \citep[e.g.,][]{Ji2003,netzel2012}. 

        Differing from previous studies, in which the failed eruptions are attributed to the erupting MFR entering in the torus stable domain after being lifted for a while \citep{liu2008,guo2010}, our observations suggest that the external reconnection could be the more important reason, in particular for those failed eruptions with a quadrupolar configuration, in which \edit2{the torus instability would likely not occur} \citep[see also][]{Mitra2022}.

		\begin{acknowledgments}
			We appreciate the referee for his/her constructive comments that improved the manuscript significantly. \emph{Solar Orbiter} is a space mission of international collaboration between ESA and NASA, operated by ESA. The STIX instrument is an international collaboration between Switzerland, Poland, France, the Czech Republic, Germany, Austria, Ireland, and Italy. \emph{SDO} is a mission of NASA's Living With a Star (LWS) program. \emph{STEREO} is the mission in NASA’s Solar Terrestrial Probes program.  \emph{Fermi} mission is a joint venture of NASA, the United States Department of Energy, and government agencies in France, Germany, Italy, Japan, and Sweden. This work is funded by the National Key R\&D Program of China under grant 2021YFA1600504 and by NSFC under grant 12127901. 
		\end{acknowledgments}

		\bibliography{ref}
		\bibliographystyle{aasjournal}
		
		\clearpage	
		
		\clearpage
   
        \clearpage

		\clearpage

		\clearpage
  
		\clearpage
		
	\end{CJK*}
\end{document}